\documentclass[]{aa}
\usepackage{graphicx}
\usepackage{amssymb,amsmath}
\usepackage{txfonts}
\usepackage{natbib}
\usepackage{epsfig}

\def\hbeta{H$\beta$}

\def\msun{M$_{\odot}$~}

\def\lsigma{$L-\sigma$~}

\def\lbeta{L(H$\beta$)~}

\def\wbeta{EW(H$\beta$)~}

\begin{document}

\title{HII galaxies as standard candles: evolutionary corrections \thanks{Partly based on data products from observations made with ESO Telescopes at the La Silla Paranal Observatory under ESO programme ID 179.A-2006 and on data products produced by the Cambridge Astronomy Survey Unit on behalf of the VHS and VIDEO consortia.} }

\author { Jorge~Melnick\inst{1,2}   and Eduardo~Telles\inst{2}}

\institute {European Southern Observatory, Av. Alonso de Cordova 3107, Santiago, Chile,
\and
Observatorio Nacional, Rua Jos\'e Cristino 77, 20921-400 Rio de Janeiro, Brasil
}

\offprints{Jorge Melnick \email{jmelnick@eso.org}}
\date{}

\authorrunning{Melnick and Telles}

\titlerunning{The \lsigma relation revisited}


\abstract{Over the past decade the relation between the Balmer-line luminosity of HII galaxies and the velocity width of the emission lines, the \lsigma relation, has been painstakingly calibrated as a cosmological distance indicator with seemingly spectacular results: the Hubble constant and the energy-density of dark energy obtained using the \lsigma indicator agree remarkably well with the values from canonical indicators.  Since most of the luminosity of these young compact starburst galaxies is emitted by a few narrow emission-lines, they can be observed with good precision up redshifts $z\sim7$ with JWST, making the \lsigma indicator a potentially unique cosmological probe. However, the precision of the method remains too low to effectively constrain the relevant cosmological parameters, notably the equation of state of dark energy. The scatter of the \lsigma relation is significantly larger than the random observational errors so we do not have a good handle on the systematics of the method. In a previous paper we posited that since the ionising radiation of these young galaxies fades rapidly over time-scales of only a few million years, age differences could be the main underlying cause of the scatter. In this paper we explore several different ways to explain the scatter of the correlation, but without success. We show that the majority of HII galaxies are powered by multiple starbursts of slightly different ages, and therefore that the equivalent widths are not reliable chronometers to correct the luminosities for evolution. Thus, it is not likely that the accuracy of the \lsigma distance indicator can be improved in the near future. Since we do not fully understand neither the systematics nor the underlying physics of the \lsigma relation, using large samples of distant HII galaxies may or may not improve the accuracy of the method.}


\maketitle

\section{Introduction}

HII galaxies are dwarf compact galaxies whose luminosities are dominated by one or more young starburst components. A significant fraction of their luminosity is emitted
in a few strong and narrow emission lines that make HII galaxies relatively easy to observe out to redshifts of cosmological relevance. HII galaxies exhibit a correlation between the luminosity of the Balmer emission lines (tipically \hbeta; \lbeta) and the width of these lines and also of the stronger [OIII] lines.  

This correlation (the \lsigma relation) has been calibrated as a distance indicator using Giant HII Regions in local late-type galaxies with accurately measured distances to define the zero point. (\citealt{GonzalezMoran2021} and references therein). 
 
The main advantage of the \lsigma distance indicator is that it can be applied to galaxies out to $z\sim7$ \cite{Chavez2024}; The main drawback is that the scatter of the relation is not only substantially larger than the observational errors, but also not well understood. So we really do not have a good handle on the systematic errors associated with the cosmological parameters derived using HII galaxies, which may be quite significant given that the dispersion of distance moduli for high-z galaxies derived from \lsigma is of several magnitudes \citep{GonzalezMoran2021, Chavez2024} 

Although the ages of the starburst clusters that power HII galaxies are selected to be shorter than a few million years (Myr), in this short age span the ionising radiation emitted by these clusters evolve significantly. So from this evolution alone we expect a scatter at least similar, if not larger, than that observed.

In our study of 2234 high-excitation HII galaxies from the Sloan Digital Sky Survey (SDSS), we showed that the scatter in the relation between \hbeta\ luminosity, and the mass of the ionising (starburst) cluster determined from SED models, correlates very well with the equivalent width of \hbeta\ (\wbeta) as expected from {\it Starburst99} models \citep{Telles2018}: Galaxies with equivalent widths larger than the average of the sample are systematically above the regression line, and vice-versa.
  
The same result does not obtain in the sample of low-redshift HII galaxies used in the calibration of the \lsigma relation \citep{Melnick2017}.  We suggested in that paper that
the problem could be due to contamination of the continuum by evolved (old and intermediate-age) stars through the \wbeta.  In this paper we pursue that idea using SED population synthesis models to disentangle the effect of evolved stars.

\section{The \lsigma relation revisited}

In \cite{Melnick2017} we explored the scatter of the relation using new velocity dispersions determined from the [OIII]5007 lines rather than the traditional H$\beta$, and photometric data from the SDSS instead of the ad-hoc wide-aperture spectrophotometry from \cite{Chavez2014}. The rationale behind these choices, as explained in detail in \cite{Melnick2017}, are: a different set of systematics; an improved S/N both in the photometry and in the line-widths; a better match between the photometric and spectroscopic apertures; and, most importantly, significantly more accurate equivalent widths of H$\beta$, which are necessary to correct the luminosities for evolutionary effects.

The HII galaxies used to define and calibrate the \lsigma relation are selected to be very young by including only objects with large equivalent widths (EW(H$\beta$)$>$50\AA), which  from Starburst99 (SB99) models \citep{Leitherer1999}, correspond to ages younger than 5-6 million years (Myr). Still, these ages are comparable to the main-sequence life times of the ionising stars making it necessary to correct the luminosities for evolutionary effects.  

\begin{figure}[ht]
\hspace*{-0.7cm}\includegraphics[width=0.50\textwidth]{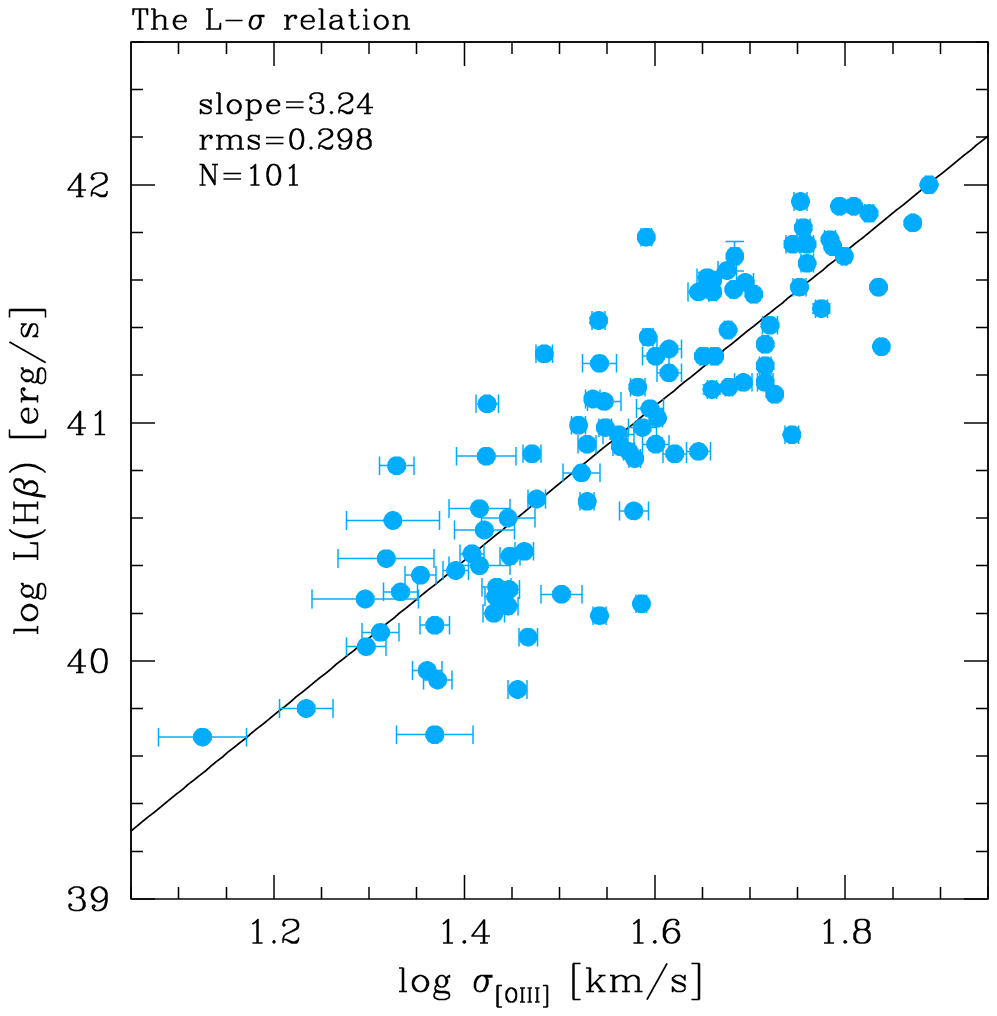} 
\vspace*{-0.8cm}
\caption{\small The observed Green \lsigma relation for HII galaxies using [OIII]5007 velocity dispersions and SDSS H$\beta$ fluxes uncorrected for evolution}
\label{LS1}
\end{figure}

Figure~\ref{LS1} presents the \lsigma relation derived from these data.  Fitting straight lines to data with errors in both variables is complicated and we refer to \cite{Melnick2017} for an in-depth discussion. Since the scatter in the \lsigma relation is significantly larger than the photometric errors, the parameters from Maximum-Likelihood solutions are virtually identical to those of standard least-squares techniques, which we will use throughout this paper.

\begin{figure}[ht]
\hspace*{-0.0cm}\includegraphics[width=0.50\textwidth]{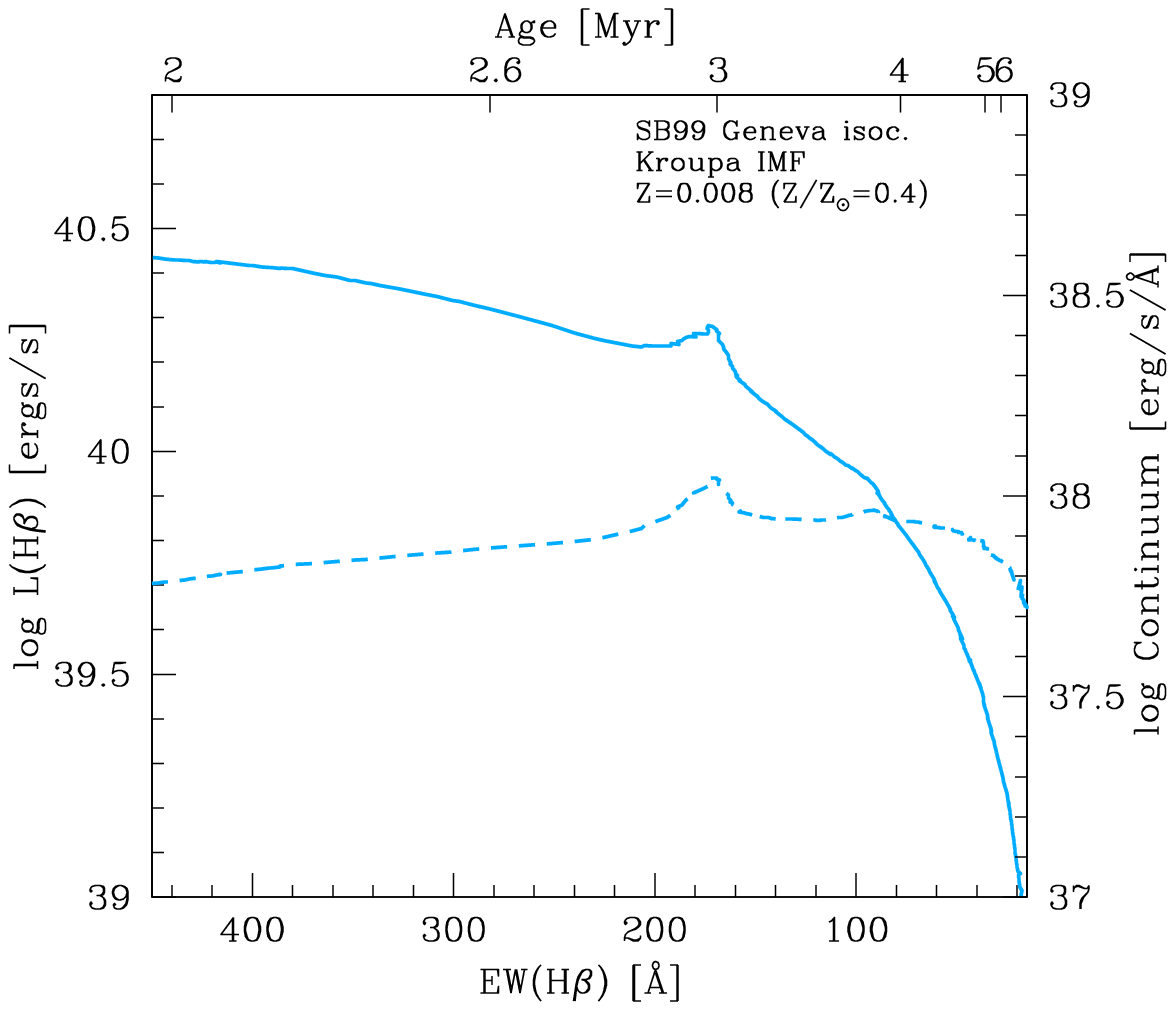} 
\vspace*{-0.6cm}
 \caption{\small Starburst99 models showing the evolution of the \hbeta\  and continuum luminosities of a single starburst as a function of the equivalent width of the \hbeta\ emission line. The solid line shows the evolution of the line luminosity and the dashed line shows the evolution of the continuum under the line.} 
\label{sb99}
\end{figure}

As mentioned above, the galaxies in our sample span a range of ages where their \hbeta\ luminosities evolve rapidly, so we expect {\it a priori} the scatter in the \lsigma relation to be correlated with age. Figure~\ref{sb99} shows single burst SB99 models using Geneva isochrones for a metallicity typical for HII galaxies of 40\% of Solar. The solid line shows the evolution of the \hbeta\ luminosity as a function of age, parametrized by the equivalent width \hbeta, \wbeta, and the dashed line shows the evolution of the continuum at the wavelength of \hbeta.

These single or "simple" models show that in the first 6~Myrs of evolution the \hbeta\ luminosity decreases by a factor of more than ten while the continuum increases by about 60\%. Most of the galaxies in our sample have \hbeta\ equivalent widths between 50 and 150$\AA$, so age differences should account for most, if not all, the scatter in the \lsigma relation. However, in \cite{Chavez2014} and \cite{Melnick2017} we find that the evolutionary corrections derived using SB99 models actually {\em increase} the scatter!

In \cite{Melnick2017} we proposed that a likely explanation for this scatter increase is that our chronometer, EW(H$\beta$), is biased by the contamination of the continuum at H$\beta$ by old and intermediate age stars. In this paper we use population synthesis models to evaluate this contamination precisely and thus correct the observed equivalent widths for contamination by evolved stars. 

\section{SED Fitting}

We use FUV to MIR broad band fluxes from GALEX; SDSS; UKIDSS; VISTA (VHS \& VIDEO); and WISE to model the SEDs. To minimise aperture matching effects we use Petrossian magnitudes except for GALEX for which we use model magnitudes. To minimise systematic effects due to different extinction laws, we first correct the fluxes for foreground Galactic extinction using the values tabulated in SDSS DR13 and a standard Galactic extinction law, and we then apply an internal extinction correction using either a Calzetti or a Gordon extinction law in a subsequent step.  

\subsection{CIGALE Models}

We used the package CIGALE \citep{Noll2009}, which offers substantial flexibility for dealing with the various complexities presented by HII galaxies. Full details of our modelling are presented \citep{Telles2018}.  For the present investigation we used our Model2 results where we feed CIGALE with fluxes corrected for internal extinction using the Balmer decrements (H$\alpha$/H$\beta$) and the reddening law of \cite{Gordon2003} appropriate to the bar of the SMC and the 30 Doradus region in the LMC, but also allowing CIGALE to make additional corrections (which were almost never necessary or very small). In Model2 we include the WISE W1 (3.7$\mu$) and W2 ($4.5\mu$) bands, but not W3 and W4 where dust-emission dominates the fluxes. 

We used a special star-formation history module kindly provided by M. Bocquien to fit three single bursts: young, intermediate, and old, of a range of ages and durations that were modelled using BC03 SSP for a Chabrier IMF. Since the nebular continuum is very relevant, particularly in the IR, we used the nebular module of CIGALE leaving the ionising parameter (logU) as a free parameter (-1.5; -2.0; or -3.0). We found that including the WISE W1 and W2 bands increased the stability of the solutions, although at these wavelengths dust emission begins to be important, thus requiring to include a dust emission module. We found  best results for the dl2014 \citep{dl2007} module, which is the one we used in Model2. The metallicity was allowed to be either Z=0.0040 or Z=0.0080, whichever gives the best fit.  

\begin{figure*}[ht!]
\includegraphics[width=0.488\textwidth]{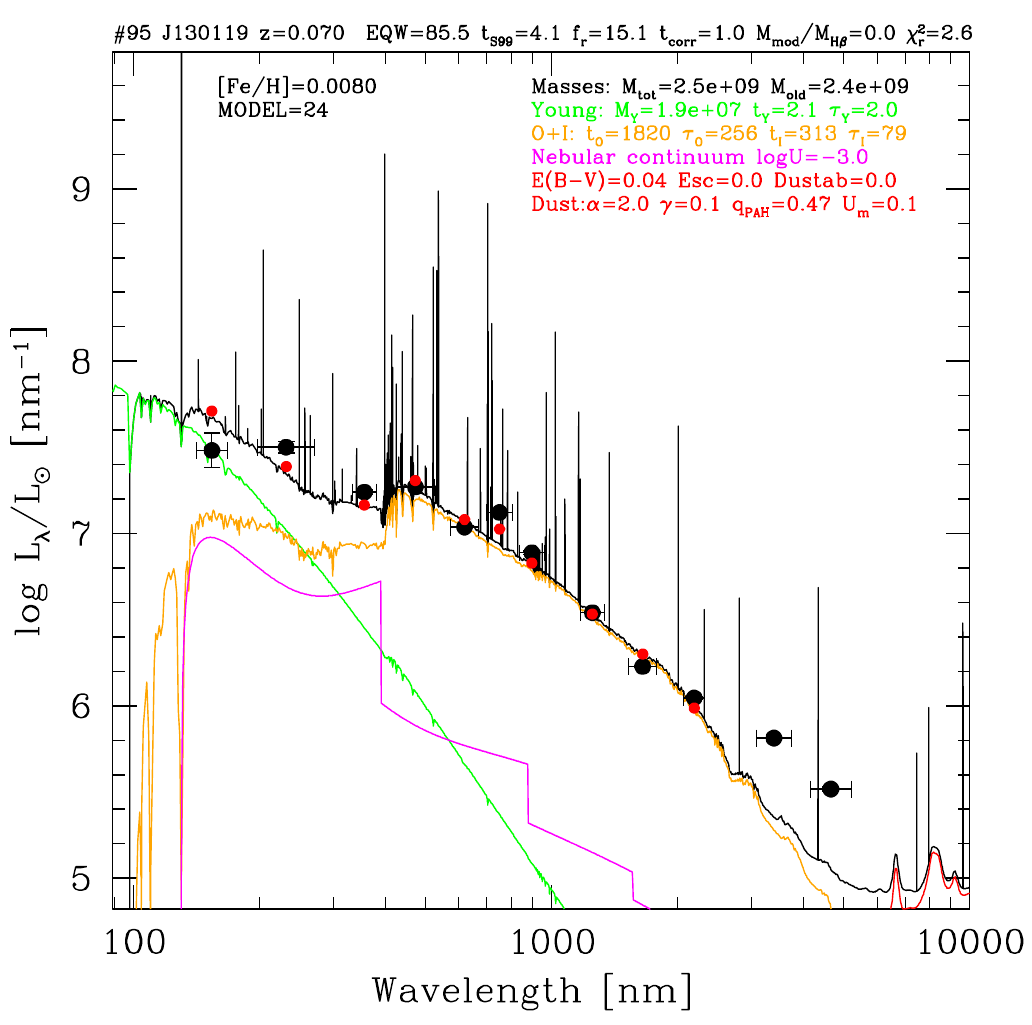}\hspace*{0.3cm}\includegraphics[width=0.49\textwidth]{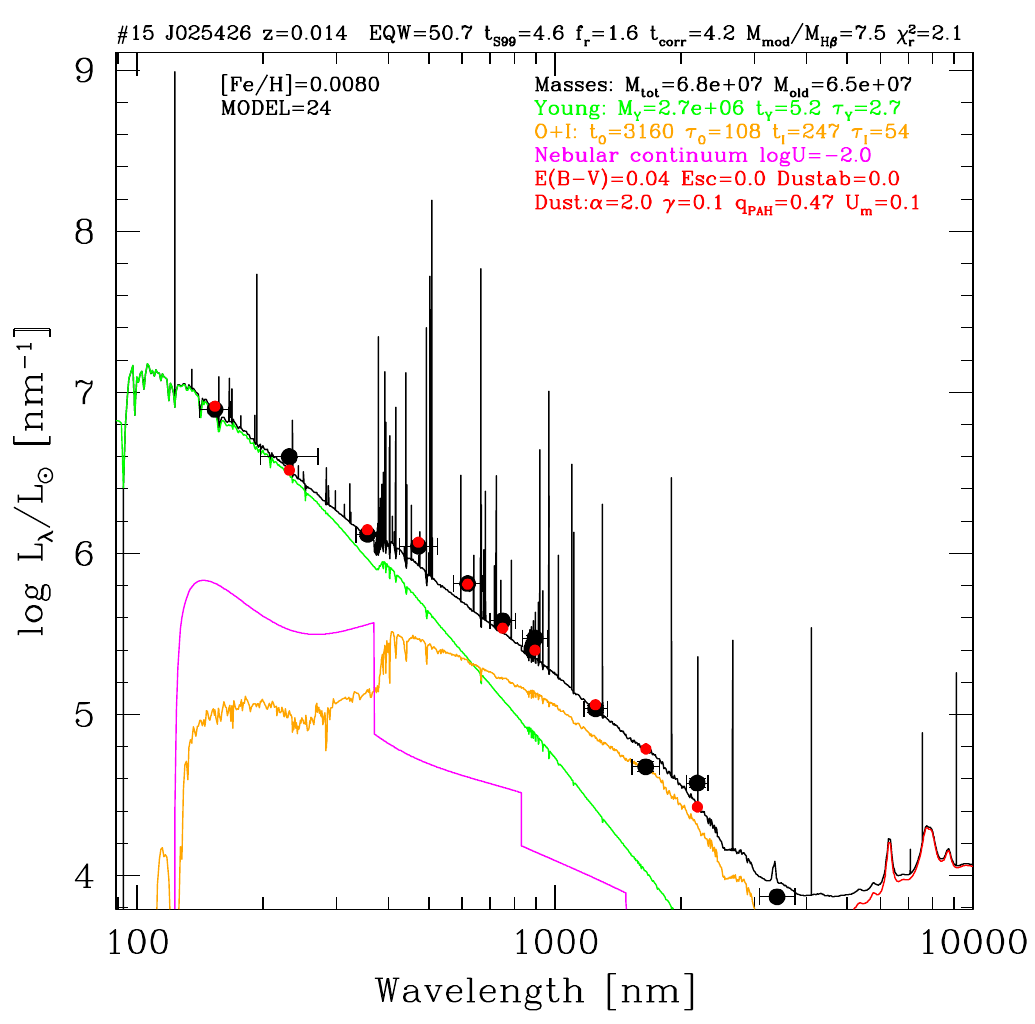} 
\vspace*{-0.1cm}
\caption{\small {\bf Left}. CIGALE model for J130119. The black line is the best fit from GIGALE to the observations shown by the black dots. The red dots are the best CIGALE fits convolved with the relevant filter bandpasses. The green line is the young stellar component fit and the orange line the old+intermediate age component. The red line is the dust emission model from \protect{\cite{dl2007}} for the parameters indicated in the figure legend; and the magenta line shows the nebular continuum. 
In this case evolved stars dominated the continuum at most bands, so $f_r=15.1$. {\bf Right}. Same as left, but for J025426 where the young stellar population dominates the continuum at \hbeta\ thus $f_r=1.6$. The WISE W3 and W4 bands are shown for reference but were not used for the fits (no red dots). The top legend on both plots show several parameters including the reduced $\chi^2$ of the solution. More information about these two galaxies is provided in the text.}
\label{example}
\end{figure*}

Examples of CIGALE fits for two HII galaxies in our sample are presented in Figure~\ref{example}.

From the best fitting spectral decomposition provided by CIGALE (young and old+intermediate age) it is straightforward to deduce the correction to the observed equivalent widths due to contamination by old and intermediate-age stars, which we parametrise as,
 \begin{equation}
f_r= \rm \frac{EW(H\beta)_{Young}}{EW(H\beta)_{observed}}
\end{equation}

\begin{figure*}[ht!]
\includegraphics[width=0.50\textwidth]{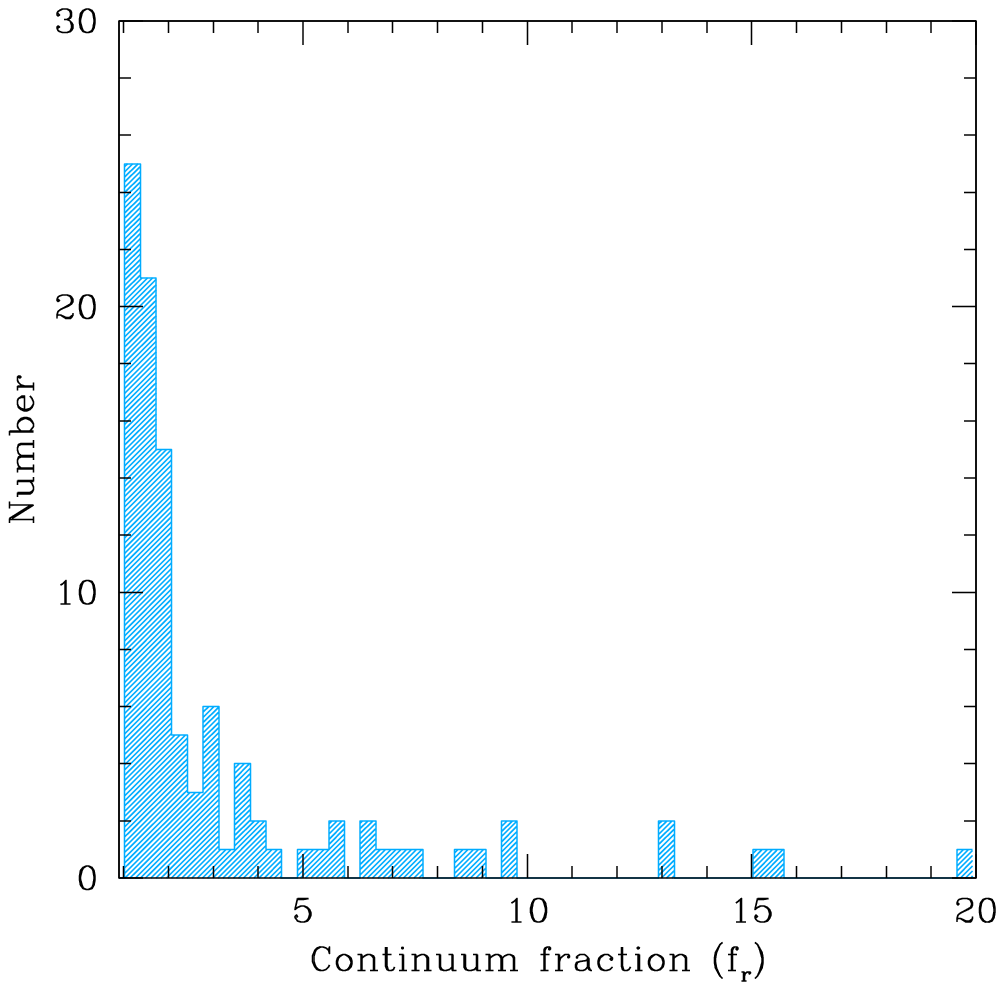}\includegraphics[width=0.50\textwidth]{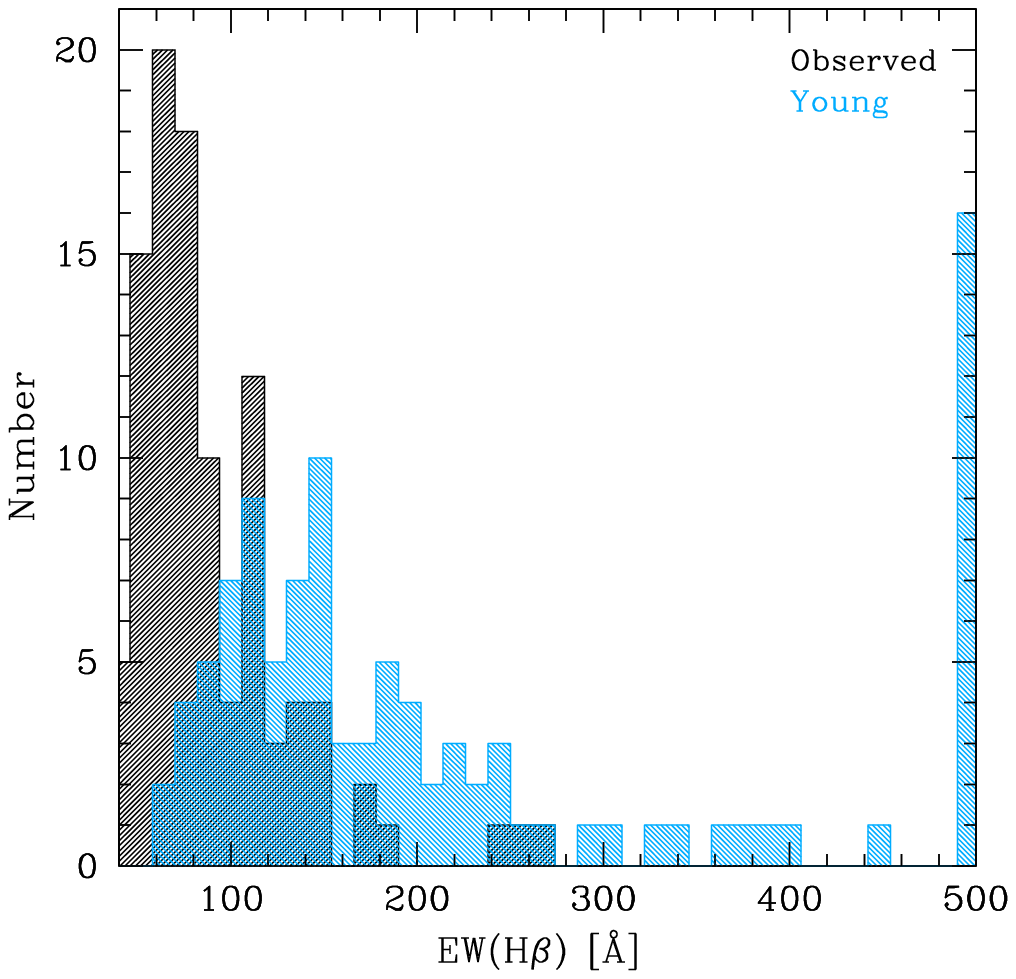} 
\vspace*{-1.2cm}
\caption{\small {\bf Left}. Distribution of $f_r$ values that measure the contamination of the continuum at H$\beta$ by evolved stars. The distribution peaks at $f_r = 1.1-1.2$ indicating that while all the HII galaxies in this sample have significant underlying old and intermediate-age stellar populations, for most galaxies the effect in EW(H$\beta$) is seen to be below 50\%.
{\bf Right}. Histograms of the distribution of observed EW(H$\beta$) in black, and corrected (Young) in blue using SED fitting. The distribution of ages [EW(H$\beta$)] of the young stellar populations is seen to be much flatter than the distribution of observed equivalent widths.}
\label{FR}
\end{figure*}

The distribution of $f_r$ values for our sample is presented in the left-hand plot of Figure~\ref{FR}, and the observed and the corrected (young) distribution of equivalent widths are shown in the right panel. The corrected distribution of ages is now quite different from the observed one.
 
\subsection{The L-$\sigma$ relation revisited}

As in \citealt{Melnick2017} hereafter Paper~I, we use the velocity dispersions from the [OIII]5007 lines, and the \hbeta\ fluxes measured on SDSS spectra to examine the \lsigma relation. The extinction corrections are derived from the Balmer decrement (H$\alpha$/H{$\beta$) without corrections for underlying absorption.  The correlation is presented in Figure~\ref{LS1} and is the same we called the "mixed" \lsigma relation in that paper.
 
Table~\ref{TB0} presents the least-squares fit parameters of the \lsigma relation with and without corrections for evolution using the equivalent widths as chronometers.
The table shows  that, contrary to our original expectations, the CIGALE corrections actually {\em increase} the scatter of the \lsigma relation quite substantially.  Thus, either our estimations of the "true" equivalent widths are completely off, or other parameters are involved (or both!). 

\begin{figure}[ht!]
\hspace*{-0.7cm}\includegraphics[width=0.50\textwidth]{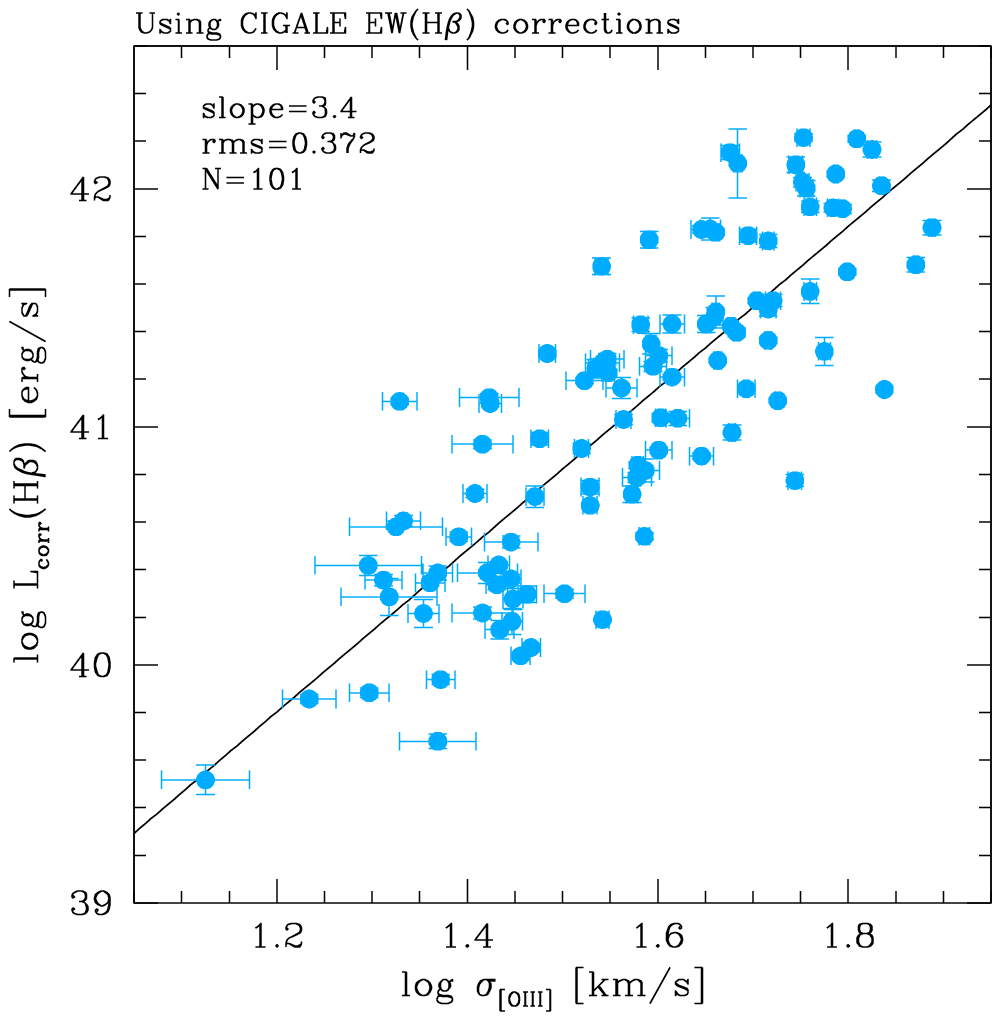} 
\vspace*{-0.8cm}
\caption{\small The \lsigma relation using the luminosities corrected for evolution using the equivalent widths of H$\beta$ corrected for contamination by evolved stars using our CIGALE models.}
\label{LS2}
\end{figure}

\begin{table}[h]
\caption{\bf Evolutionary Corrections for the full sample}
\tabcolsep 2.0mm
\tiny
\begin{tabular}{l l l l}  
\hline\hline 			
			& Without age 		& Corrected using  			& Corrected using  		\\ 
			& corrections		& observed EW(H$\beta$)	& CIGALE EW(H$\beta$)	\\ \hline
slope		& $3.24\pm0.18$	&  $3.50\pm0.21$			& $3.40\pm0.23$		\\
rms		  	& $0.298$			&  $0.337$				& $0.372$				\\  \hline
 \end{tabular}
\label{TB0}
\end{table}

\section{Deconstructing the scatter of the \lsigma relation}
 
We have seen that {\it naive} evolutionary corrections using \wbeta\ as an evolutionary clock do not explain the scatter of the \lsigma relation, and that refining our  clock through modelling of the underlying stellar continuum does not help. This is puzzling (and disappointing) because the evolutionary corrections are substantial and should stand out over other effects.

\subsection{Multiple starbursts}

A possible way out of this conundrum is that HII galaxies are actually made of multiple starbursts of different ages. For example, from Figure~\ref{sb99} it is easy to see that the sum of two starbursts of the same mass, but of different ages, say 2~Myr and 5~Myr, has an \hbeta\  luminosity only 10\% larger than the 2~Myr old burst alone, but half the equivalent width. So, if HII galaxies are made of multiple starbursts of different ages, their integrated \hbeta\ equivalent widths may not be a reliable evolutionary clock. 

Following \cite{Telles1997}, \cite{Fernandez2018} performed a morphological classification of the HII galaxies in our sample dividing them in two groups: type~I for galaxies with disturbed or irregular isophotes; and type~II for galaxies with regular isophotes, which were also found to be more compact.

We found that 54 of 103 galaxies in the sample of \cite{Fernandez2018}  are morphological singles (type~II) of which 47 are in our sample.   Figure~\ref{LS53} shows the observed \lsigma relation for these 47 galaxies; the parameters of the relation corrected for evolution using the equivalent widths of the starburst component are shown in Table~\ref{TBis}. 

\begin{figure}[ht!]
\hspace*{-0.7cm}\includegraphics[width=0.5\textwidth]{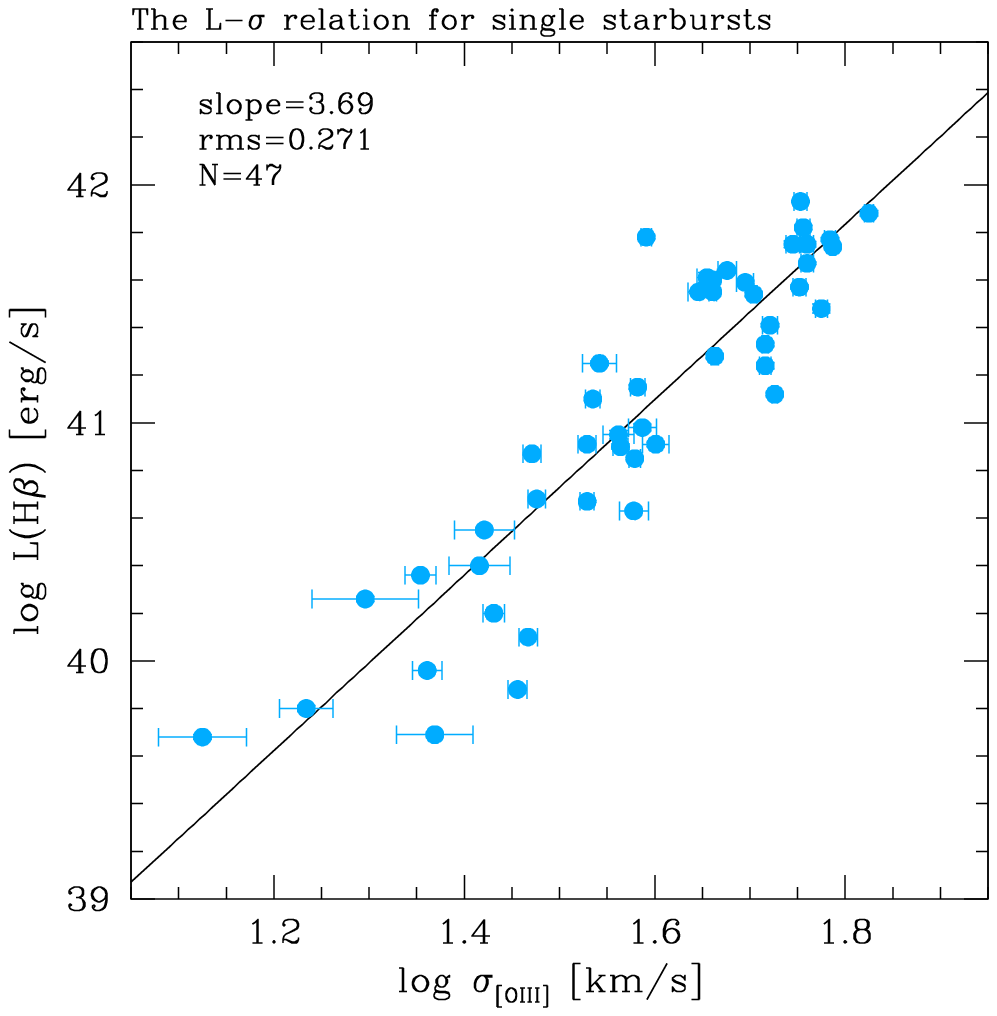}
\vspace{-0.8cm}
\caption{\small The \lsigma relation for HII galaxies classified as single starbursts on the basis of morphology.} 
\label{LS53}
\end{figure}

\begin{table}[ht]
\caption{\bf Evolutionary corrections for starbursts of single morphology.}
\tabcolsep 2.0mm
\tiny
\begin{tabular}{l l l l}  
\hline\hline 			
			& Without age 		& Corrected using  			& Corrected using  		\\ 
			& corrections		& observed EW(H$\beta$)	& CIGALE EW(H$\beta$)	\\ \hline
slope		& $3.68\pm0.25$	&  $4.03\pm0.27$			& $4.11\pm0.30$		\\
rms		  	& $0.271$			&  $0.298$				& $0.326$				\\  \hline
 \end{tabular}
\label{TBis}
\end{table}

Table~\ref{TBis}  shows that correcting for evolution using the equivalent widths of \hbeta\ with or without corrections for contamination by evolved stars does not reduce the scatter of the \lsigma relation for single starbursts. On the contrary, as was the case for the full sample, these corrections actually increase the scatter quite significantly! 

Notice, however, that not all HII galaxies classified as single on the basis of optical morphology have Gaussian [OIII] profiles. In fact, selecting only single galaxies with Gaussian profiles (determined by visual inspection) leaves only 22 galaxies, to few to allow further analysis.

  \subsection{Other parameters}

Of the various observables tested by \cite{Chavez2014} (and later \cite{Fernandez2018}) as possible additional parameters of the \lsigma relation, the size of the objects as measured by the Petrossian radii in u'-band SDSS images ($R_u$), gave the strongest signal. $R_u$ was chosen because contamination by nebular lines and continuum is minimal in the $u'$ band. However, as shown in Figure~\ref{example}, contamination by the nebular continuum and/or by evolved stars is dominant for several objects in our sample.
   
Table~\ref{TB1} shows the results of repeating the analysis of \citep{Chavez2014} using the radii of \cite{Fernandez2018}, but our velocity dispersions and luminosities and the corresponding evolutionary corrections.  

The table shows the results of fitting,
\begin{equation}
\label{rer}
 log L(H\beta) = c_0 +c_{1}\times log \sigma_{[OIII]} + c_2\times log R_u
\end{equation}
using standard LSQ techniques. 

\begin{table}[h]
\caption{\bf Radius as a second parameter}
\tabcolsep 2.0mm
\tiny
\begin{tabular}{l l l l}  
\hline\hline 			
			& Without age 		& Fluxes corrected using  		& Fluxes corrected using  		\\ 
Parameter		& corrections		& observed EW(H$\beta$)		& CIGALE EW(H$\beta$)		\\ \hline
$c_0	$		& $34.96\pm0.25$	&  $34.70\pm0.27$			& $33.84\pm0.29$			\\
$c_1$	  	& $2.12\pm0.21$	&  $2.11\pm0.22$			& $2.42\pm0.24$			\\
$c_2	$ 		& $1.00\pm0.12$	& $1.21\pm0.13$			& $1.24\pm 0.14$			\\ 
rms			& 0.223			& 0.242					& 0.261					\\ \hline
 \end{tabular}
\label{TB1}
\end{table}
Table~\ref{TB1} shows that including the radius as a second parameter leads, at face value, to a significant reduction of the scatter, and that the \lsigma relation appears actually to be a correlation between the integrated \hbeta\ luminosity of HII galaxies and $R_u\sigma^2$, reminiscent of some sort of dynamical mass.  This can be visualized in Figure~\ref{lulu}, where the \hbeta\ luminosity is plotted against this "dynamical" mass, $M_{dyn}=R_u\sigma^2$

\begin{figure}[ht]
\hspace*{-0.7cm}\includegraphics[width=0.50\textwidth]{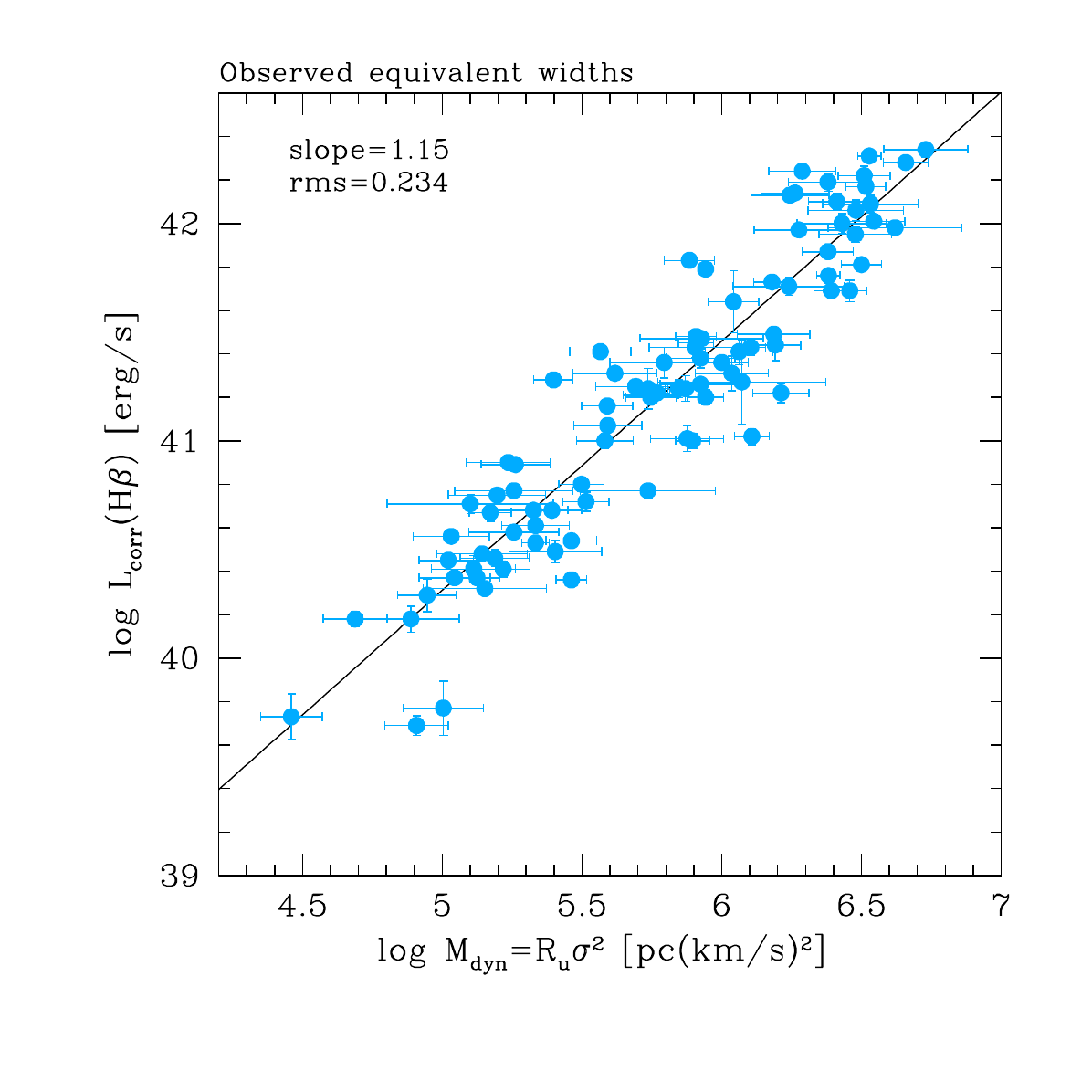}
\vspace*{-0.8cm}
\caption{\small Relation between \hbeta\ luminosity corrected for evolution using the observed equivalent widths, and the "dynamical" mass defined as $M_{dyn}=R_u\sigma^2$ in the text. The line shows a least-squares fit of parameters shown in the figure labels.}
\label{lulu}
\end{figure}

This correlation brings us back to the arcane discussion of whether HII galaxies (and their local siblings the Giant HII Regions) are self-gravitating balls of ionised gas (see \citealt{Terlevich2018} for a recent discussion and a comprehensive list of references).  We know that the velocity dispersions of Giant HII Regions (GHR), as measured by the width of their emission lines, are not due to the motions of ionised blobs of gas moving in the gravitational potential of these systems, but rather the result of the superposition of multiple stellar-wind (actually cluster-wind) driven bubbles of ionised gas  \citep{Melnick2021}. Since GHR and HII galaxies follow the same \lsigma relation, it would be very surprising if their velocity dispersions had a different physical origin, which would also invalidate the use of GHR as zero point calibrators of the \lsigma distance indicator.  

Our strong view is that in both classes of objects the gas motions are driven by stellar winds and that the tight correlation between \lbeta and  $R_u\sigma^2$ is not due to gravity but has a different explanation: Malmquist bias. 

Figure~\ref{malok} plots  $\sigma$ as a function of distance $D$. The least-squares fit shown by the line has the functional form $\sigma \sim D^{0.39}$, so $R_u\sigma^2$ goes as $D^{1.8}$, and in log-log we expect a tight correlation between \lbeta  (or M$_{young}$) and $R_u\sigma^2$  of slope close to unity as we observe. The ranges spanned by the \hbeta\ fluxes and angular sizes of HII galaxies being much smaller than their range of distances. 

\begin{figure}[ht]
\hspace*{-0.7cm}\includegraphics[width=0.50\textwidth]{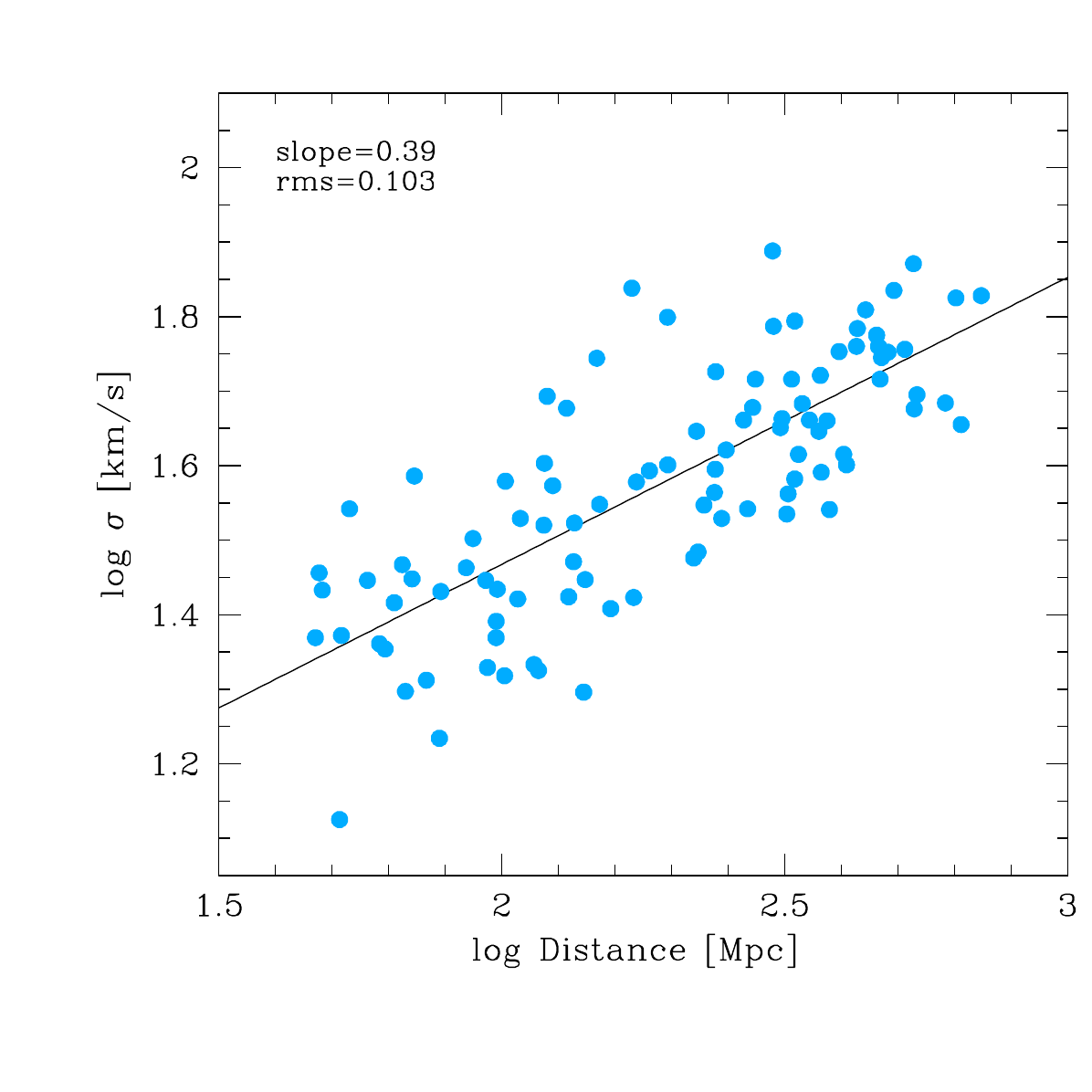}
\vspace*{-0.8cm}
\caption{\small Malmquist bias in our flux-limited sample of HII galaxies produces a correlation between the distance independent parameter $\sigma$ and distance. } 
\label{malok}
\end{figure}

Every time we deal with log-log plots of slope very close to unity  where some parameter (in our case the distance) appears in both axes to a similar power, we must ask ourselves whether we have a true correlation, or a "Stork" diagram. \footnote{The logarithm of the number of Storks per capita in European cities correlates perfectly with the logarithm of the number of births per capita; Storks bring babies!}  

It is easy to show by scrambling the (angular) radii, or by simply replacing $R_u$ by the distance, that Figure~\ref{lulu} is a classical "Stork" diagram: With our present sample, it is not possible to establish whether there is any correlation between luminosity and starburst dimensions, even if such a correlation could be envisaged if the sizes correlate with cluster mass.

Moreover, our CIGALE models show that the contribution of nebular emission and evolved stars to the light in the SDSS $u'$ band varies significantly, while about 80$\%$ of the galaxies in our sample have multiple components. In practice, therefore, $R_u$ measures something different from one galaxy to the next. 

This supports our strong conclusion that HII galaxies are not self-gravitating balls of ionised gas and that the supersonic velocities inferred by the widths of the emission lines are due to a superposition of a large number of wind-driven bubbles of gas.
 
\section {Discussion}
\subsection{ Is \wbeta a reasonable evolutionary chronometer?}

We have seen that the observed \wbeta works better as an evolutionary chronometer than the equivalent width corrected for contamination by evolved stars from SED fits, although none of the two is adequate to explain the scatter of the \lsigma relation. We argued that the uncorrected \wbeta works better because the majority of the HII galaxies in our sample contain multiple starburst components, but we have not demonstrated that our clock actually works.

From our previous work modelling a sample of 2234 high-excitation SDSS HII galaxies \citep{Telles2018}, we know that the observed \wbeta work reasonably well in that sample, a sub-set of the parent population of 4200 SDSS high excitation HII galaxies with \wbeta>50\AA, used in \cite{Chavez2014} to select the 101 galaxies in our \lsigma sample. The number is reduced to 2234 selecting objects with published UV to MIR photometry.

In \cite{Telles2018} we found that the mass of young stars M$_{young}$ derived from SED fitting with CIGALE, correlates extremely well with  \hbeta\ luminosity.  Of course, this is a classical log-log "Stork" diagram (the square of the distance appears on both axis), but the interesting result is that the scatter is reduced to just the observational errors when the luminosities are corrected for evolution using the observed \wbeta and SB99 models.

\begin{figure*}[htt]
\includegraphics[width=0.50\textwidth]{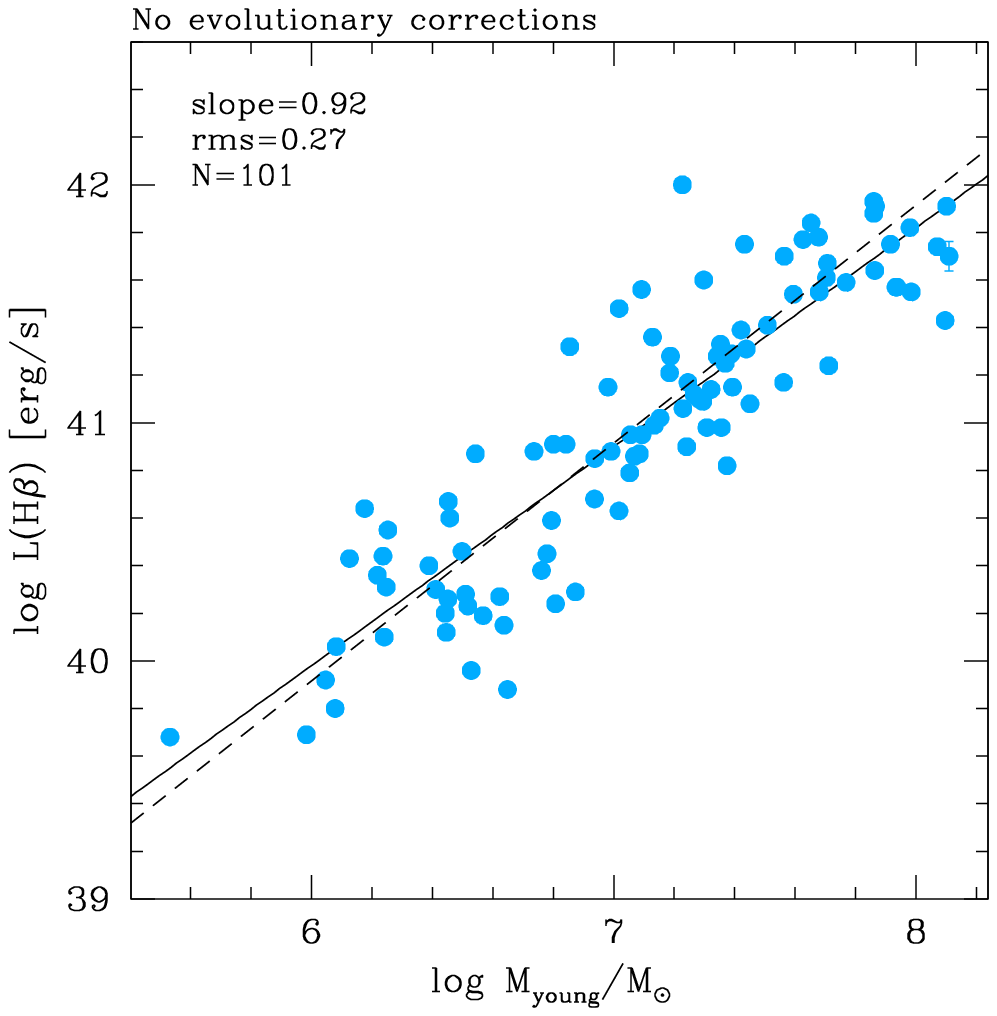}
\includegraphics[width=0.50\textwidth]{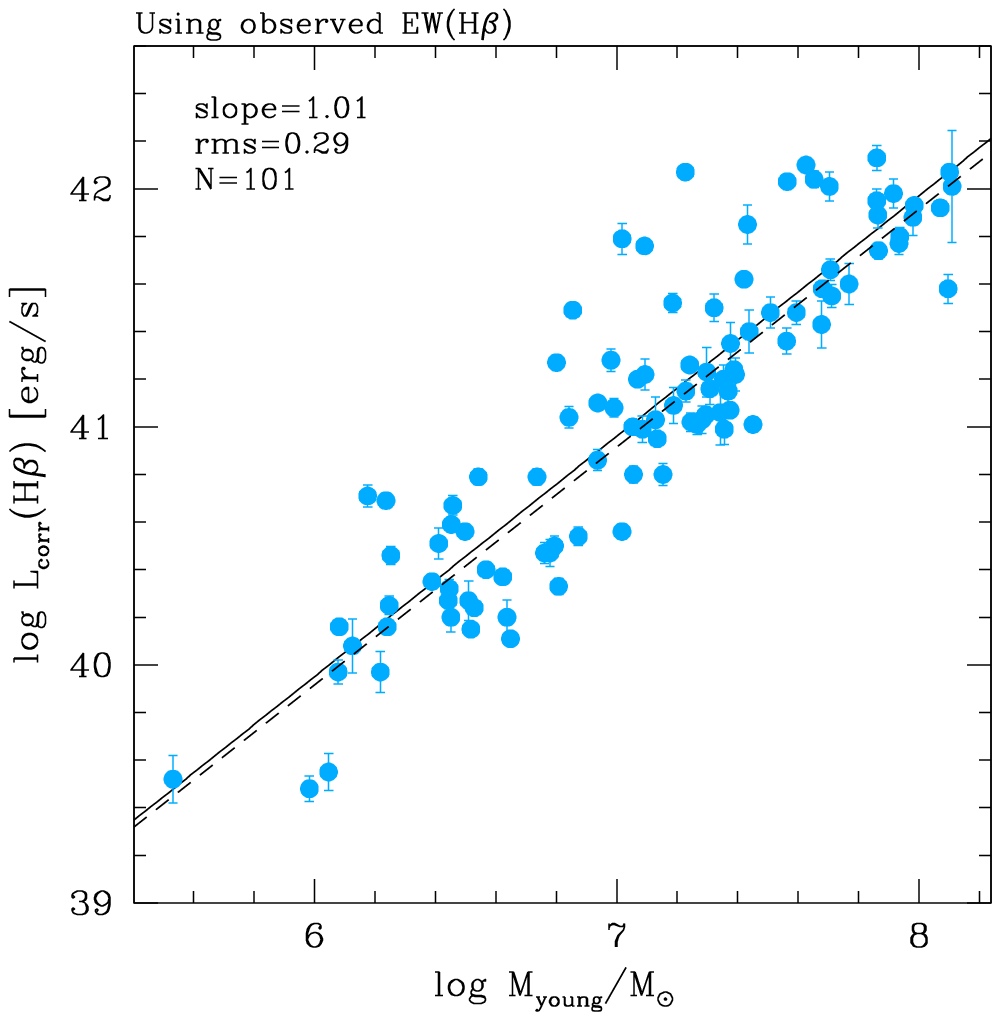}
\vspace*{-1.2cm}
\caption{\small {\bf Left}: Relation between photometric mass of young stars (M$_{young}$) from CIGALE SED fits and the observed H$\beta$ luminosity not corrected for evolution. {\bf Right}: Same as left, but with the luminosities corrected for evolution using the observed \wbeta. The solid lines shows a least-squares fits to the data; the dashed line plots the predictions of SB99 models.  As discussed in the text, this is a classical "Stork" diagram: the logarithm of the distance appears in both axes.}
\label{maza}
\end{figure*}

Figure~\ref{maza} shows the relation between M$_{young}$ and \lbeta  for the 101 HII galaxies in our [OIII]  \lsigma sample.  In the left-hand panel the luminosities are uncorrected for evolution.
 The right-hand panel shows the luminosities corrected for evolution as in \cite{Telles2018}. Although at first glance the correlation appears to be tighter that the panel on the left, the scatter is actually (slightly) worse. The dashed line plots the luminosities predicted by SB99 models (at the mean age of the sample) for the M$_{young}$ masses from CIGALE. The excellent consistency with the observed luminosities (solid line) shows that it is probably OK to use SB99 models in the present context, and that we can trust our CIGALE masses (but see below). 
  
But if that is the case, why are the \wbeta corrections not working as well as they did for the "parent" sample?  Closer inspection of the right panel shows that the scatter is dominated by just a few objects. In fact there are 10 objects with deviations of more than a factor of 3 from the best fit line that have a common feature: their luminosities are dominated by the old+intermediate age component, not by the young stars. Figure~\ref{example} shows the SED of one of these objects, J130119.

So the answer to the title of this section is: yes, but the \lsigma sample contains many objects with SEDs dominated by old+intermediate age stars that require large corrections. Removing such objects improves the indicator substantially but at a high cost: removing luminous HII galaxies (that contain multiple starbursts and tend to have low equivalent widths) makes it more difficult to find suitable HII galaxies at high redshifts.

\subsection{Does the turbulence of the ionised gas evolve together with \lbeta?}

The turbulence of the ionised gas could be evolving together with the luminosity, due, for example, to the onset of Type-Ic supernovae.   Figure~\ref{sisi} plots the velocity dispersion $\sigma$ as a function of the mass of young stars from the CIGALE models. The points are color-coded according to CIGALE continuum correction factor $f_r$. 

The correlation between $\sigma$ and M$_{young}$ is weaker thant expected. Part of the scatter is due to the outliers discussed in the previous section, but not all. Most of the outliers in this plot do not show any identifiable peculiarities.  Incidentally, the $\sigma$--M$_{young}$ relation is also noisy in the "parent" sample, but with a significantly flatter slope ($0.118\pm0.008$ versus $0.214\pm0.017$ shown in the figure). Malmquist bias may explain the difference in slope, but we have not investigated this any further.

So the answer to the title of this section is: No. With the exception of the "outliers" we do not see any correlation between the scatter in the relation between $\sigma$ and M$_{young}$ and \wbeta. 

\begin{figure}[ht]
\hspace*{-0.7cm}\includegraphics[width=0.5\textwidth]{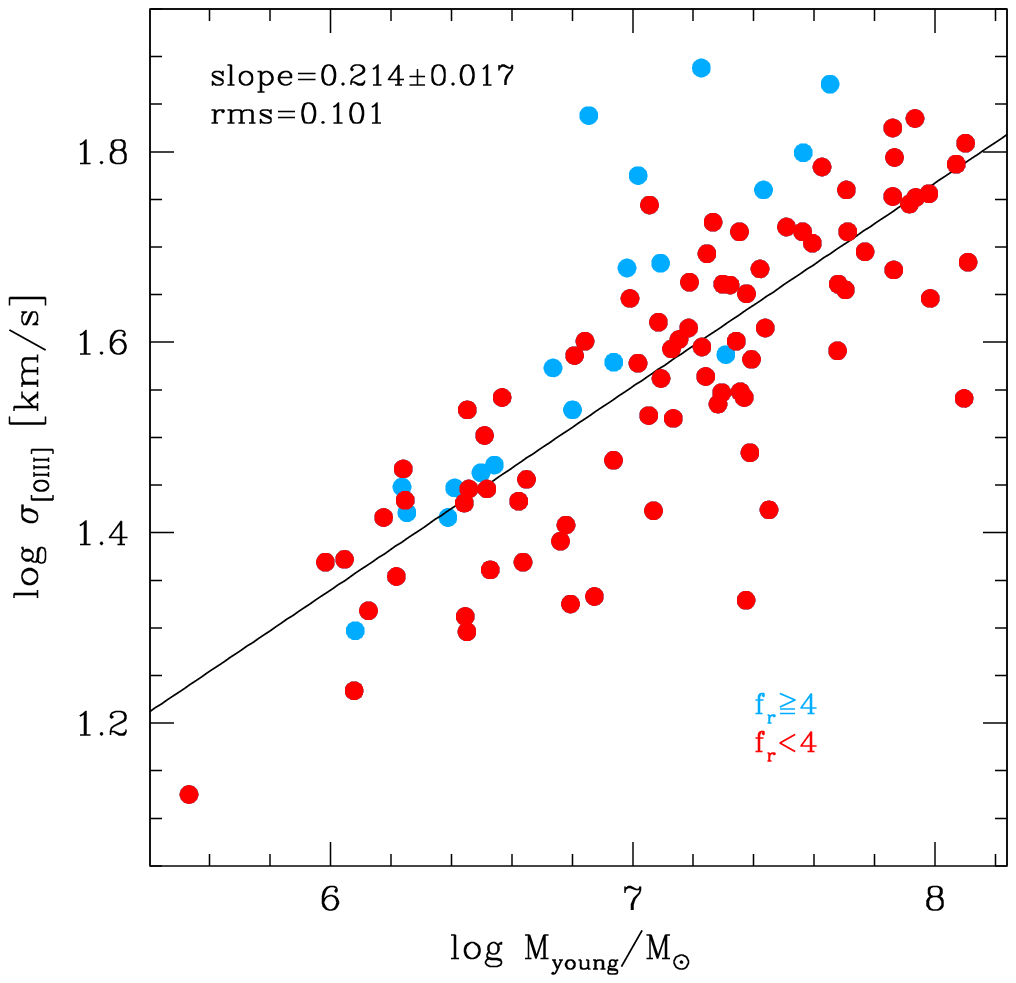}\
\vspace*{-0.8cm}
\caption{\small Relation between photometric mass of young stars (M$_{young}$) from CIGALE SED fits and the equivalent width of \hbeta. The colours code objects with dominant old+intermediate stellar populations that have large continuum corrections $f_r$.}
 \label{sisi}
\end{figure}

\section {Conclusion: the scatter of the \lsigma relation}

Several factors determine the scatter of the relation,

\begin{enumerate}
   \item Outliers. A few peculiar objects have an inordinate effect on the scatter. The Optical-IR fluxes of these objects are dominated by Old+Intermediate stars, whereas by definition the SED's of HII galaxies should be dominated by the young stellar component;
   \item Evolution. The range of ages spanned by the galaxies in the \lsigma sample imply that evolutionary corrections to the \hbeta\ luminosities are important. However,
   about 80\% of the HII galaxies in our sample appear to have multiple starburst components, either as seen on direct images or from multiple bumps in the integrated [OIII] profiles. Multiplicity reduces the efficacy of \wbeta as an evolutionary chronometer so, although we do detect the effects of evolution, with the available tools and observations it is not possible to normalise the luminosities to some fiducial age;
   \item  The turbulence of the ionised gas as measured by the widths of the emission lines is only loosely correlated with the mass of the ionising clusters. The velocity dispersion scales as $\sigma\sim M^{1/5}_{young}$, indicating that the transformation of stellar winds into the {\em cluster winds} that ultimately inflate the bubbles of ionised gas is a highly dissipative process where most of the kinetic energy of the stellar winds is thermalised at high temperatures \citep{Melnick2021}. This suggests that the mass and the age of the ionising stars may not be the only parameters the govern the physics of the \lsigma relation;
   \item Given that observed the scatter of the relation is not very sensitive to evolutionary corrections, which should be large, it is unlikely that subtler effects, such as changes in the IMF, or the metallicity of the stars, contribute significantly to the the scatter;
   \item Observations of the starburst cluster 30~Doradus  (see \cite{Crowther2019} for a recent review) indicate that HII galaxies may contain tens to hundreds of super massive stars ($M>150$\msun), which are not included in either SB99 or CIGALE models. Our analysis of the evolution of HII galaxies, therefore, may be incomplete, in particular for the youngest objects.

\end{enumerate}

With the presently available sample, it does not appear possible to reduce the scatter of the \lsigma relation below $\sim0.3$dex. It may be possible to improve the precision of the relation as a distance indicator by selecting galaxies with substantially larger equivalent widths.  As it stands, the \lsigma relation does not seem to be precise enough to be used as a cosmological distance indicator. Not all HII galaxies can be used as standard candles.

\section*{Acknowledgements}
We are grateful to  Denis Burgarella, the father of CIGALE, and to Mederic Bocquien for guiding us through our first steps with the code and answering numerous questions. Mederic kindly wrote the special module to fit three stellar populations that we used in this work. JM acknowledges support from a CNPq {\it Ci\^encia sem Fronteiras} grant at the Observatorio Nacional in Rio de Janeiro, and the hospitality of ON as a PVE visitor.

Funding for the SDSS and SDSS-II has been provided by the Alfred P. Sloan Foundation, the Participating Institutions, the National Science Foundation, the U.S. Department of Energy, the National Aeronautics and Space Administration, the Japanese Monbukagakusho, the Max Planck Society, and the Higher Education Funding Council for England. The SDSS Web Site is http://www.sdss.org/.

The SDSS is managed by the Astrophysical Research Consortium for the Participating Institutions. The Participating Institutions 
are the American Museum of Natural History, Astrophysical Institute Potsdam, University of Basel, University of Cambridge, 
Case Western Reserve University, University of Chicago, Drexel University, Fermilab, the Institute for Advanced Study, the 
Japan Participation Group, Johns Hopkins University, the Joint Institute for Nuclear Astrophysics, the Kavli Institute for Particle
Astrophysics and Cosmology, the Korean Scientist Group, the Chinese Academy of Sciences (LAMOST), Los Alamos National
Laboratory, the Max-Planck-Institute for Astronomy (MPIA), the Max-Planck-Institute for Astrophysics (MPA), New Mexico State 
University, Ohio State University, University of Pittsburgh, University of Portsmouth, Princeton University, the United States 
Naval Observatory, and the University of Washington.

The entire GALEX Team gratefully acknowledges NASA's support for construction, operation, and science 
analysis for the GALEX mission, developed in corporation with the Centre National d'Etudes Spatiales of 
France and the Korean Ministry of Science and Technology. We acknowledge the dedicated team of engineers,
technicians, and administrative staff from JPL/Caltech, Orbital Sciences Corporation, University of California, 
Berkeley, Laboratoire d'Astrophysique Marseille, and the other institutions who made this mission possible. 
 
The UKIDSS project is defined in Lawrence et al. (2007). UKIDSS uses the UKIRT Wide 
Field Camera WFCAM (Casali et al. 2007).  The photometric system is described in 
Hewett et al. (2006), and the calibration is described in Hodgkin et al. (2008). The 
pipeline processing and science archive are described in Irwin et al. (2009, in prep) 
and Hambly et al (2008).

We gratefully acknowledge use of data from the ESO Public Survey programme ID 179.B-2004 taken with the VISTA telescope, data products from CASU and VSA archive operated by WFAU.

This publication makes use of data products from the Wide-field Infrared Survey Explorer, which is a joint project of the University 
of California, Los Angeles, and the Jet Propulsion Laboratory/California Institute of Technology, funded by the National 
Aeronautics and Space Administration

\bibliographystyle{aa}
\bibliography{mybib}
\label{lastpage}

\end{document}